\documentclass[twocolumn,nofootinbib,
showpacs,prl,aps,floatfix]{revtex4}
\usepackage{mathtext}
\usepackage{mathrsfs}
\usepackage{bm}
\usepackage{graphicx}
\begin{document}
\title{Dynamics of a Tonks-Girardeau gas released from a hard-wall trap
}
\author{A. del Campo and J. G. Muga
}
\address{Departamento de Qu\'\i mica-F\'\i sica,
Universidad del Pa\'\i s Vasco, Apdo. 644, Bilbao, Spain}%
\def\la{\langle}
\def\ra{\rangle}
\def\om{\omega}
\def\Om{\Omega}
\def\vep{\varepsilon}
\def\wh{\widehat}
\def\tr{\rm{Tr}}
\def\da{\dagger}
\newcommand{\beq}{\begin{equation}}
\newcommand{\eeq}{\end{equation}}
\newcommand{\beqa}{\begin{eqnarray}}
\newcommand{\eeqa}{\end{eqnarray}}
\newcommand{\intf}{\int_{-\infty}^\infty}
\newcommand{\into}{\int_0^\infty}
\begin{abstract}
We  study the expansion dynamics of a Tonks-Girardeau gas released from a hard wall 
trap. Using the Fermi-Bose map, the density profile is found analytically 
and shown to differ from that one of a classical gas in 
the microcanonical ensemble even at macroscopic level, for any observation 
time larger than a critical time. The relevant time scale arises as a 
consequence of fermionization.

\end{abstract}
%
%
\maketitle
%
%
%
%
%
%
%
%
%
The Tonks-Girardeau (TG) regime \cite{Girardeau60} is that of 1D impenetrable bosons, 
which is relevant to atom waveguide experiments with low densites and temperatures
 and large scattering lengths 
\cite{Olshanii98}.
Under such conditions the radial degrees of freedom are reduced to 
zero-point oscillations, 
resulting a 1D effective system, as it has been demonstrated in several 
experiments \cite{exp}.
Some remarkable studies dealing with dynamics in this regime 
have shown the limits of 
mean-field theory in splitting-recombination processes \cite{GW05bis}, 
the spatial focusing of the probability density through Talbot 
oscillations \cite{RCB99}, and existence of dark and grey solitons in a 
toroidal trap \cite{GW00}.
During the 1D expansion of a harmonically confined Tonks gas, 
fermionization of the system was observed in the momentum 
distribution \cite{MG05,OS02}. A parallel study pointing out the ``reciprocal'' 
character of the Fermi-Bose duality, was recently carried out in 
\cite{GM05} for a fermionic TG gas which undergoes a dynamical 
bosonization.
In the mean time, and motivated by the experimental build-up of square 
well \cite{HHHR01} and 
hard wall optical box traps \cite{MSHCR05}  a growing interest has been developed 
concerning low dimensional Bose gases trapped in such geometries \cite{Cazalilla02, BGOL05}
after the seminal paper by Gaudin \cite{Gaudin71}.
However, most of the studies have dealt with the gas within the trap and, at variance 
with the case of a harmonic confinement, only the single-particle evolution has 
been considered, in the field of ultracold neutron interferometry \cite{GK76} 
and diffraction in time \cite{Godoy02}.

In this Letter we account for a detailed study of the expansion dynamics of 
a TG gas, after switching off the confining hard wall potential and show the 
deviation from the associated classical gas in the microcanonical ensemble.
In such a regime the Fermi-Bose (FB) map \cite{Girardeau60,YG05,CS99,GW00b} gives 
the many body wavefunction of $N$ strongly interacting bosons from the one of a 
free Fermi gas with all spins frozen in the same direction.
In order to do so, it suffices to apply the ``antisymmetric unit function" 
$\mathcal{A}=\prod_{1\leq j<k\leq N}sgn(x_{k}-x_{j})$, as $\psi_{B}(x_{1},\dots,x_{N})=
\mathcal{A}(x_{1},\dots,x_{N})\psi_{F}(x_{1},\dots,x_{N}).$
The Fermi wavefunction, antisymmetric under permutation of particles,
 is built as a Slater determinant, with one particle in each eigenstate 
of the trap
\beq
\psi_{F}(x_{1},\dots,x_{N})=\frac{1}{\sqrt{N!}}det_{n,k=1}^{N}\phi_{n}(x_{k}) .
\eeq
One more advantage of the general FB mapping is that it holds for time dependent 
processes (governed by one-body external potentials), 
since the $\mathcal{A}$ operator does not include time explicitly
\cite{GW00b,YG05}, this is, Fermi statistics holds under time evolution.
The glaring upshot is that as far as local 
coordinate distributions are concerned, 
to deal with the manybody Tonks gas it suffices to work out the single 
particle problem, since $\vert\psi_{B}(x_{1},\dots,x_{N};t)\vert^{2}
=\vert\psi_{F}(x_{1},\dots,x_{N};t)\vert^{2}$.
%
%
%
%
%
In particular, from the involutivity of the $\mathcal{A}$ operator ($\mathcal{A}^{2}=1$) 
and the fact that $\la\phi_{n}\vert U^{\dag}U\vert\phi_{m}\ra=\delta_{nm}$, where $U$ is 
the time-evolution operator, 
it follows that the time-dependent density profile can be calculated as \cite{GW00b}
\beqa
\label{dp}
\rho(x,t)&=&N\!\!\int\vert\psi_{B}(x,x_{2},\dots,x_{N};t)\vert^{2}dx_{2}
\cdots dx_{N}\nonumber\\
&=&\sum_{n=1}^{N}\vert\phi_{n}(x,t)\vert^{2}.
\eeqa
%

%

%
%
%
%
%
{\it Single eigenmode dynamics.}
Motivated by the Eq. (\ref{dp}), 
in this section we tackle the problem 
of studying the time evolution of the $n$-th eigenstate of a hard wall trap.
As it is well-known they have the general form 
$\phi_{n}(x,t=0)=\sqrt{\frac{2}{L}}\sin(n\pi x/L)\chi_{[0,L]}(x)$,
%
with $n\in\mathbf{N}$ 
where the characteristic function 
can be conveniently written as a difference of Heaviside functions, 
$\chi_{[0,L]}(x)=\Theta(L-x)-\Theta(-x)$.
Concerning the dynamics we consider the free evolution under the 
propagator
\beq
K(x,t\vert x',0)=\sqrt{\frac{m}{2\pi i\hbar t}}
e^{\frac{im(x-x')^{2}}{2t\hbar}}
\eeq
after suddenly switching off the trap at zero time.
Using the superposition principle and introducing 
$p_{n}=\hbar n\pi/L$,
\beqa
\label{phit}
\phi_{n}(x,t) & = &\int_{-\infty}^{\infty}dx' 
K(x,t\vert x',t=0)\psi(x',t'=0)\nonumber\\
& = &\frac{1}{2i}\sqrt{\frac{2}{L}}\sum_{\alpha=\pm}\alpha
\big[e^{i\alpha p_{n}L/\hbar}M(x-L,\alpha p_{n}/\hbar,\hbar t/m)\nonumber\\
&  &-M(x,\alpha p_{n}/\hbar,\hbar t/m)\big]
\eeqa
with

\beqa
\label{Moshfunc}
& & M(x,p/\hbar,\hbar t/m) := \int_{-\infty}^{0}dx'K(x,t\vert x',t'=0)
e^{ipx'/\hbar}\nonumber\\
& & =\frac{e^{i\frac{mx^{2}}{2t\hbar}}}{2}w\bigg[-\frac{1+i}{2}\sqrt{\frac{t}{m\hbar}}
\left(p-\frac{mx}{t}\right)\bigg],
\eeqa
and $w(z)=e^{z^{2}}erfc(-iz)$ the so called Faddeyeva function 
\cite{Faddeyeva}.
After \cite{Moshinsky52}, $M(x,k,\tau)$ has been named 
the Moshinsky function. Similar expressions have been 
derived in the field of ultracold 
neutron interferometry \cite{GK76}, 
and discussed recently in the context of 
diffraction in time \cite{Godoy02}.
\begin{figure}
\includegraphics[height=5cm]{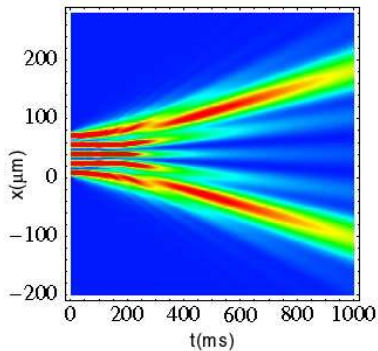}
\caption{\label{mode} Density plot of the probability 
density for the dynamical evolution for the fifth eigenstate (${}^{87}$Rb atom,
$L=80\mu m$).}
\end{figure}
When the particle is trapped in the $n$-th mode, and at short times after being released, 
the probability density presents $n$ maxima, but the central ones 
tend to fade away with time as shown 
in Fig. \ref{mode}.
Indeed, the general structure of the $n$-th eigenstate ($n>1$) under evolution presents 
a bifurcation in two main branches after the semiclassical time 
$t_{n}=m L/(2 p_{n})=mL^{2}/(2n\pi\hbar)$.
%
%
%
%

\begin{figure}
\includegraphics[height=5cm]{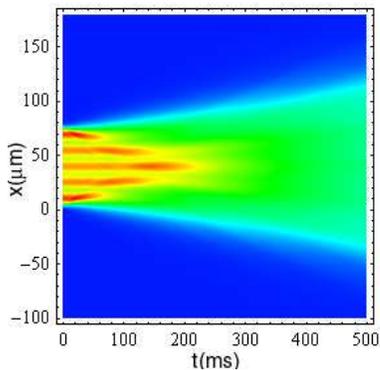}
\caption{\label{figdp} Density plot of the probability 
density $\rho$ for a Tonks gas composed of $N=5$ atoms of ${}^{87}$Rb ($L=80\mu m$).}
\end{figure}

{\it Tonks gas dynamics.}
%
%
%
Remarkably, as stated above, the calculation  of the density profile is 
possible from  Eqs. (\ref{dp}),(\ref{phit}), and (\ref{Moshfunc}).
As illustrated in Fig. \ref{figdp}, the density profile exhibits at short times a peaked 
structure  with the number of maxima equal to that of particles.

This spatial antibunching results from the underlying 
fermionization characteristic of the Tonks regime. 
Indeed, the two-particle local correlation, 
$g_{2}=\la\hat{\Psi}^{\dag}(x)^{2}\hat{\Psi}(x)^{2}\ra
/\la\hat{\Psi}^{\dag}(x)\hat{\Psi}(x)\ra^{2}$, 
was shown to tend to zero
 even for non-zero temperatures \cite{KGDS03}.
However, the visibility of the interference pattern is lost both with 
increasing time and the number of bosons under consideration.

\begin{figure}
\includegraphics[height=7cm,angle=-90]{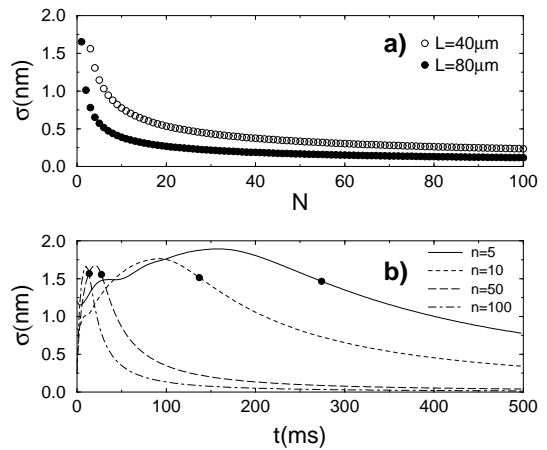}

\caption{\label{ntonks} a) Dependence of the variance on the number of ${}^{87}$Rb atoms 
for the density profile of TG gas confined in a trap.
b) Time dependence of the variance of the density 
profile for different number of ${}^{87}$Rb atoms. 
Filled circles mark the corresponding $t_{N}$ and $L=80\mu m$.
}
\end{figure}

A measure of the roughness of the density profile is the root mean 
square of the density profile weighted with itself, $\sigma_{\rho_{N}}
=\sqrt{\la\rho_{N}\ra_{\rho_{N}}^{2}-\la\rho_{N}^{2}\ra_{\rho_{N}}}$, 
with $\rho_{N}$ the density profile normalised to one particle.
In fact, at time equals zero and as a function of the particle number 
this measure decreases monotonically as shown in Fig. \ref{ntonks}a, 
in agreement with the resolution of the identity within the box, 
$lim_{N\rightarrow\infty}\rho={\bf 1}_{[0,L]}$, see Eq. (\ref{dp}).
The time evolution is richer in structure, presenting an initial positive slope, 
see Fig. \ref{ntonks}b, 
to finally start to decrease after $t\sim t_{N}$, reaching zero asymptotically.

\begin{figure}
\includegraphics[height=5cm,angle=-90]{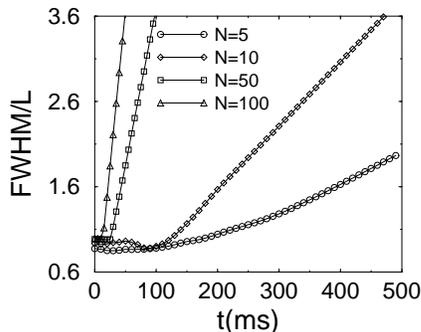}
\caption{\label{tonksw} Time dependence of the FWHM of a cloud of ${}^{87}$Rb atoms
released from a trap $80\mu m$ in width.}
\end{figure}

After a transient regime, where the structure of the peaks varies in time, 
the roughness of the density profile becomes monotonically decreasing 
with both time and number of particles.
%
%
%
In particular, the fastest components will be associated with the highest 
excited state (the Fermi level in the dual system of non-interacting spin-frozen 
fermions), and therefore with quasi-momentum given by 
$\pm p_{N}=\pm\hbar N\pi/L$.
Actually, these components govern the width of the expanding 
cloud for $t\gtrsim t_{N}$.
Figure \ref{tonksw} shows the variation in time of the full width at 
half maximum (FWHM) for different number of particles.
The upshot is that deviations from a linear dependence on time are observed right after 
switching off the trap, similarly to the results reported in \cite{OS02} for 
the harmonically confined Tonks gas. Such deviations disappear with increasing number 
of particles and already for $N\sim100$ are bellow the millisecond time scale. 
From its definition, one  can conclude that $t_{N}$, 
the time necessary for a classical particle 
to leave the trap when it is initially located in the center of the box and moves 
with the momentum $p_{N}$, arises as a consequence of 
fermionization.
Nevertheless, it is clear from the transient features of the single-particle 
solution that a self-similar expansion does not occur for a TG gas released 
from a hard wall trap. This fact contrasts with the harmonic case, 
pointing out the relevance of the confining geometry.

{\it Comparison with a classical gas.}
Next we consider a classical gas of distinguishable particles interacting through 
a contact potential.
Since for the TG gas 
the total energy ($E=\sum_{n=1}^{N}E_{n}, E_{n}=\hbar^{2}n^{2}\pi^{2}/(2mL^{2})$)
is fixed once the number of particles $N$ in the initial trap is specified \cite{BGOL05}, 
it is natural to deal with a microcanonical ensemble in the classical limit.
This leads to a distribution in phase space of the form:
\beq
\rho_{cl}({\bf x},{\bf p},0;E) =\mathcal{N}_{0}
\delta\left(E-\sum_{i=1}^{N}\frac{p_{i}^{2}}{2m}\right)
\prod_{i=1}^{N}\Theta(x_{i})\Theta(L-x_{i}),
\eeq
where the normalization reads 
$\mathcal{N}_{0}=N\Gamma(N/2)/(2m\pi^{N/2}L^{N}\varrho^{N-1})$, 
$\Gamma(z)$ is the Gamma function and $\varrho=\sqrt{2mE}$.
Using Liouville's theorem, momentum conservation 
(but for interchange), and the symmetry under permutation 
of particles in the microcanonical ensemble, one finds
$\rho_{cl}({\bf x},{\bf p},t;E)
=\rho_{cl}({\bf x}-{\bf p}t/m,{\bf p},0;E)$, where 
${\bf x}=\{x_{1},\cdots,x_{N}\}$, and ${\bf p}=\{p_{1},\cdots,p_{N}\}$.
To obtain the density profile, an integration over 
$\{x_{2},\dots,x_{N},p_{1},\dots,p_{N}\}$ variables has to be performed. 
Note that each of the integrals over $x_{i}$ contributes exactly $L$.
Moreover, all the integrals over momenta can be carried out by using 
the generalisation of spherical polar coordinates to a hypersphere 
in a $N$-dimensional space subjected to the constraint 
$\sum_{i}p_{i}^{2}=\varrho^{2}=2mE$ \cite{Sommerfeld49,MW95}, which fixes 
the radius $\varrho$ of the hypershell.
Since it is always possible to choose one coordinate of the form 
$p=p_{1}=\varrho\cos\xi$, it follows that
\beqa
\rho_{cl}(x,t;E) &=& \mathcal{N}\int_{0}^{\pi}
d\xi \sin^{N-2}\xi\Theta(x-t\varrho\cos\xi/m)\nonumber\\
& & \times\Theta(L+t\varrho\cos\xi/m-x),
\eeqa
where $\mathcal{N}=N\Gamma(N/2)/(\sqrt{\pi}L\Gamma[(N-1)/2])$ 
is the normalisation constant.



\begin{figure}
\includegraphics[height=5cm,angle=-90]{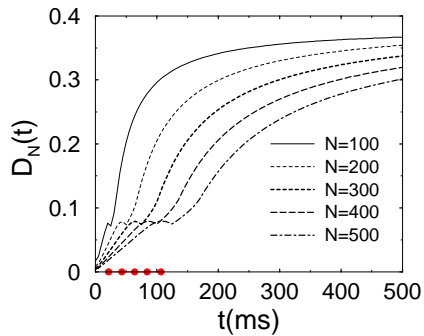}
\caption{\label{dqcl} Measure $D_{N}$ for a fixed density,
$d_{0}=1$  atom$/\mu m$
and increasing number of particles (${}^{87}$Rb). Filled circles mark $t_{N}$ 
and verify $D_{N}(t_{N})\simeq 0.08$.} 
\end{figure}

For a systematic comparison between the Tonks 
and microcanonical gas, we introduce the time-dependent measure 
$D_{N}(t)=\int dx \vert\rho(x,t)-\rho_{cl}(x,t)\vert/N$, 
which is plotted in Fig. \ref{dqcl} for different number of particles.
As a result, a clear time scale given by $t_{N}$ can be established, such that 
for small times ($t/t_{N}<1$) both quantum and classical profiles essentially 
coincide ($D_{N}<0.1$). 

\begin{figure}
\includegraphics[height=7cm,angle=-90]{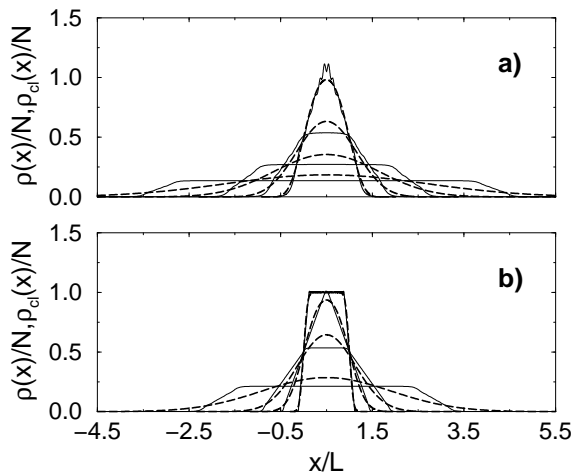}
\caption{Microcanonical (dashed lines) and quantum 
(continuous line) density profiles 100ms after expansion for: 
a) fixed length box $L=80\mu m$, and 
 $N(t_{N})$=10 (137ms), 25 (55ms), 50 (27ms), and 100 (1ms), from top to bottom. 
b) fixed density $d_{0}=1$atom$/\mu m$, and $N(t_{N})$
=100 (21ms), 250 (54ms), 500 (107ms), and 2000 (428ms),  from bottom to top.}
\label{thlim}
\end{figure}

In particular, the microcanonical model reproduces  
for short times and low number 
of particles the quantum profile in a coarse-grained fashion, 
neglecting inteference (see Fig.\ref{thlim}a). It is precisely 
for low $N$ when the differences between the initial quantum spatial 
distribution and the uniform classical one, are greater.
For larger number of particles the quantum profile tends to ``resolve the identity''
 within the trap, reaching a uniform constant distribution when 
$N\rightarrow\infty$. 
This is, the quantum result is a staggeringly flat distribution but for the 
smoothed edges, and can be understood as a consequence of the discrete spectrum 
of the free Hamiltonian confined in the subspace $[0,L]$, 
with equally spaced quasimomenta $p_{n}$ (See Fig.\ref{mode} ).
Indeed, in the limit $t/t_{N}\ll1$ and $N\gg1$, the profile is well described by the
characteristic function of the form 
$\Theta(x+p_{F}t/m)\Theta(L+p_{F}t/m-x)/(L+2p_{F}t/m)$ (Fermi-hat), 
which expands with the momentum of the Fermi level in the dual system, 
$p_{N}$.
Similar profiles have already been reported for an effective 1D
weakly interacting fermionic gas confined in a harmonic trap \cite{ID05}.

By contrast, it can be proved that the momentum distribution 
of the microcanonical gas is of the form \cite{MW95}
\beq
f_{cl}^{(N)}(p)=\frac{\Gamma(N/2)}
{\sqrt{\pi \varrho^{2}}\Gamma((N-1)/2)}
\left(1-\frac{p^{2}}{\varrho^{2}}\right)^{(N-3)/2}
\eeq  
which in the reduced variable $p/\varrho$, for large $N$, tends to be a 
Gaussian with zero mean and $1/N$ variance.
It follows that the density profile becomes also Gaussian in this limit, 
in clear disagreement  with the TG profile, as shown in Fig. 
\ref{dqcl} for $t>t_{N}$. 

 

In the thermodynamic limit, achieved at a given observation time and 
fixed density $d_{0}=N/L$ by taking $N\rightarrow\infty$,
 both the quantum and microcanonical density profiles 
become indistinguishable of one another. 
Since for fixed density $t_{N}= mN/(2\pi\hbar d_{0}^{2})$, 
the agreement is reached more easily at short times after 
switching off the trap.
Figure \ref{thlim}b shows that at an observation 
time of $100\mu s$, 2000 particles suffice.
Deviation from the classical gas increases with time 
which acts as a lense pointing out the difference in the 
underlying momentum distributions. 
Indeed, taking note of the experimental number of atoms in \cite{MSHCR05}, 
generally between 500-3500, a rectangular profile is to be expected in the 
TG regime both in the trap and after switching it off, 
being overall described by the Fermi-hat model. 
\newline
{\it Discussion.}
Taking advantage of the Fermi-Bose map, the dynamics 
of a strongly interacting many-body system can be  worked out 
once the time evolution of the single-particle eigenstates of 
the confining hard-wall trap is known.
In this way, we have accounted for the first study of a 
many-body Moshinsky shutter problem 
\cite{Moshinsky52,DM05,GCR97} including interactions.
Right after switching off the trap, the profile evolves through a 
transient regime where the overall behaviour is analogous to the one of 
a classical gas in the microcanonical ensemble. 

However, at times $t>t_{N}=mL^{2}/(2N\pi\hbar)$ 
and in the same conditions $(N,L)$, 
a regime with uniform density is observed in the Tonks gas, 
clearly deviating from the classical bell-shaped profile.
The origin of this main feature can be traced back to %
 the discrete spectrum of the original trap potential. 
Finally, it is noteworthy that the results obtained in this Letter 
equally hold for a noninteracting Fermi gas 
due to the involutivity of the antisymmetric unit function which 
entails the 1D Fermi-Bose duality.

\begin{acknowledgments}
This paper has benefited from inspiring comments by 
David Gu{\'e}ry-Odelin, Dirk Seidel and Andreas Ruschhaupt.
This work has been supported by Ministerio de Educaci\'on y Ciencia
(BFM2003-01003) and UPV-EHU (00039.310-15968/2004).
A.C. acknowledges financial support by the Basque Government (BFI04.479). 
\end{acknowledgments}
%
%





\begin{thebibliography}{10}
\expandafter\ifx\csname natexlab\endcsname\relax\def\natexlab#1{#1}\fi
\expandafter\ifx\csname bibnamefont\endcsname\relax
  \def\bibnamefont#1{#1}\fi
\expandafter\ifx\csname bibfnamefont\endcsname\relax
  \def\bibfnamefont#1{#1}\fi
\expandafter\ifx\csname citenamefont\endcsname\relax
  \def\citenamefont#1{#1}\fi
\expandafter\ifx\csname url\endcsname\relax
  \def\url#1{\texttt{#1}}\fi
\expandafter\ifx\csname urlprefix\endcsname\relax\def\urlprefix{URL }\fi
\providecommand{\bibinfo}[2]{#2}
\providecommand{\eprint}[2][]{\url{#2}}


\bibitem{Girardeau60} M. Girardeau, J. Math. Phys. {\bf 1} 516, (1960).

\bibitem{Olshanii98} M. Olshanii, Phys. Rev. Lett. {\bf 81} 938, (1998); 
V. Dunjko, V. Lorent, and M. Olshanii, 
{\it ibid} {\bf 86}, 5413 (2001).

\bibitem{exp} B. Paredes {\it et al} Nature (London) {\bf 429}, 227 (2004); 
T. Kinoshita {\it et al.}, 
Science {\bf 305}, 1125 (2004).

\bibitem{GW05bis} M. D. Girardeau and E. M. Wright, Phys. Rev. Lett. {\bf 84} 5239 (2005).

\bibitem{RCB99} A. Rojo,  J. L. Cohen, and P. R. Berman,
 Phys. Rev. A {\bf 60} 1482 (1999).

\bibitem{GW00} M. D. Girardeau and E. M. Wright, Phys. Rev. Lett. {\bf 84}, 5239 (2000).

\bibitem{MG05} A. Minguzzi and D. M. Gangardt, Phys. Rev. Lett. {\bf 94}, 240404 (2005).

\bibitem{OS02}  P. {\"O}hberg and L. Santos, Phys. Rev. Lett. {\bf 89}, 240402 (2002).

\bibitem{GM05} M. D. Girardeau and A. Minguzzi, cond-mat/0508063 (2005).

\bibitem{HHHR01} W. H{\"a}nsel {\it et al.}, 
Nature {\bf 413}, 498 (2001).

\bibitem{MSHCR05} T. P. Meyrath, F. Schreck, J. L. Hanssen, C-S. Chuu, and M. G. Raizen,
Phys. Rev. A {\bf R71}, 041604 (2005). 

\bibitem{Cazalilla02} M. A. Cazalilla, Europhys. Lett. {\bf 59}, 793 (2002); 
M. A. Cazalilla, J. Phys. B {\bf 37}, S1 (2004).

\bibitem{BGOL05} M. T. Batchelor {\it et al.}, 
J. Phys. A {\bf 38}, 7787 (2005).

\bibitem{Gaudin71} M. Gaudin, Phys. Rev. A {\bf 4}, 386 (1971).

\bibitem{GK76} A. S. Gerasimov and M. V. Kazarnovskii, Sov. Phys. JETP 
{\bf 44}, 892 (1976).

\bibitem{Godoy02} S. Godoy,  Phys. Rev. A {\bf 65}, 042111 (2002).

\bibitem{YG05} V. I. Yukalov and M. D. Girardeau, Laser Phys. Lett {\bf 2}, 375 (2005).

\bibitem{CS99} T. Cheon and T. Shigehara, Phys. Rev. Lett. {\bf 82}, 2536 (1999).

\bibitem{GW05} M. D. Girardeau and E. M. Wright, Phys. Rev. Lett. {\bf 95}, 
010406 (2005)

\bibitem{GW00b} M. D. Girardeau and E. M. Wright, Phys. Rev. Lett. {\bf 84}, 5691 (2000).


\bibitem{Faddeyeva} V. N. Faddeyeva and N. M. Terentev 
{\it Mathematical Tables: Tables of the values of 
the function $w(z)$ for complex argument}, Pergamon, New York (1961);
 A. Abramowitz and I. A. Stegun 
{\it Handbook of Mathematical Functions}, Dover, New York (1965).

\bibitem{Moshinsky52} M. Moshinsky, Phys. Rev. {\bf 84}, 525 (1951); 
M. Moshinsky, {\it ibid} {\bf 88}, 625 (1952).

\bibitem{Sommerfeld49} A. Sommerfeld, {\it Partial Differential Equations in Physics}, 
Academic, New York, p.227 (1949).

\bibitem{MW95} J. G. Muga and D. M. Wardlaw,  Phys. Rev. E {\bf 51}, 5377 (1995).
 

\bibitem{KGDS03} K. V. Kheruntsyan, D. M. Gangardt, P.D. Drummond, and G. V.Shlyapnikov,
Phys. Rev. Lett. {\bf 91}, 040403 (2003).

\bibitem{ID05} A. Imambekov and E. Demler, cond-mat/0510801.

\bibitem{DM05} A. del Campo and J. G. Muga, J. Phys. A {\bf 38}, 9803 (2005).

\bibitem{GCR97} G. Garc{\'i}a-Calder{\'o}n and A. Rubio, Phys. Rev. A {\bf 55}, 3361 (1997).

\end{thebibliography}
\end{document}